\documentclass[11pt]{article}

\usepackage{amsmath}
\usepackage{amsmath}

\begin{document}

\begin{center}

\noindent {\bf   V.M. Red'kov \\
EFFECT OF MOTION OF THE REFERENCE  FRAME ON GEOMETRICAL FORM OF
THE LINE OF MONOPOLE SINGULARITY\\[2mm]
Institute of Physics,\\
National Academy of Sciences of Belarus }

\end{center}

\begin{quotation}

The monopole singularity string,  infinite direct  line in the
rest reference frame, is considered as a rigid  physical  line,
and behavior  of its geometrical form under the Lorentz
transformations  is studied. The method to test  the form of any
rigid line or surface in relativistical terms is based on the use
of light signals emitted from the origin of the reference frame.
Influence of the motion of the reference frame on the form of
singularity line is as follows: (1)  in general case, when the
line  does not go through the origin of the rest reference frame
the line in the moving reference frame becomes a hyperbola; (2) in
the special case, when the line of singularity   goes  through the
origin of the rest reference frame,  the line in the moving
reference preserves its form only modified by  the relativistical
aberration effect.

\end{quotation}

\section{ Introduction
}

In the literature concerning  magnetic monopoles,
the problem of the monopole line of singularity,  so-called monopole string, has been
considered many times and in many different aspects -- we remind only three basic papers by P.A.M. Dirac  [1-3].
One may see two opposite viewpoints: (first) the line of  monopole singularity is fictitious,
coordinate-based,  it cannot manifest itself in  physical experiments; (second) the line of monopole
singularity  is quite real and it  can be studied in physical experiments.

Certainly, the second point of  view may be interpreted as considering  the monopole just like
a one-dimensional extended  object.  In other words, the monopole is not a point-like, and thereby it
should not be considered as  a specific elementary particle. In absence of  experimental evidence
to existence of monopoles, these two views may be taken as alternative working ideas:
each of them should be studied seriously -- theoretically and experimentally.

In the present paper, we try to  consider the monopole string as a physically real line,
remembering that the Dirac  monopole potential is not defined at the infinite half-line, whereas  the Schwinger's potential
is not well defined at the whole infinite line. Then, this singular string is changed by a rigid  physical  line, and further
one can study its geometrical properties   with the help of tools of ordinary relativistic kinematics.
We investigate how the form of this line will change in dependence of  inertial motion
of the reference frame. General   method  to solve the problem is taken from [4].
The method to test  the form of any
rigid line or surface in relativistical terms is based on the use
of light signals emitted from the origin of the reference frame.

\vspace{2mm}

Influence of the motion of the reference frame on the form of
singularity line is:

\begin{quotation}

(1)  in general case, when the
line  does not go through the origin of the rest reference frame
the line in the moving reference frame becomes a hyperbola;

\vspace{2mm}

(2) in
the special case, when the line of singularity   goes  through the
origin of the rest reference frame,  the line in the moving
reference preserves its form only modified by  the relativistical
aberration effect.

\end{quotation}

\section{ Monopole string as a set of events in the  space-time}

\hspace{5mm} Let the monopole string be at rest in the reference frame $K'$
and is oriented along the axis  $z'$ - -see Fig. 1.
Let the monopole string is seen by other  inertial observer  $K$:

\vspace{2mm}

\unitlength=0.6mm
\begin{picture}(100,50)(-55,0)
\special{em:linewidth 0.4pt} \linethickness{0.6pt}

\put(0,0){\vector(+1,0){40}}  \put(+45,-5){$x'$}
\put(0,0){\vector(0,+1){40}}  \put(-10,+40){$z'$}
%\put(0,0){\vector(-1,-1){20}} \put(-30,-20){$z$}

\put(20,0){\line(0,+1){30}} \put(20.3,0){\line(0,+1){30}}
\put(20,0){\line(0,-1){30}} \put(20.3,0){\line(0,-1){30}}

\put(80,0){\vector(+1,0){40}}  \put(+125,-5){$x$}
\put(80,0){\vector(0,+1){40}}  \put(+70,+40){$z$}
%\put(80,0){\vector(-1,-1){20}} \put(+50,-20){$z'$}

\put(+35,+30){\vector(+1,0){20}}
\put(45,+35){$\mbox{v}$}

\end{picture}

\vspace{25mm}

\begin{center}
{\bf Fig. 1}
\end{center}

In the moment of coincidence  of two reference frame ($t'_{1}=t_{1}=0$) let
from the origin light signal   be emitted in different directions (event  $1'$); which arrives at
the moment  $t_{2}'$   to a point on the line  -- (event $2'$).
This light signal in the reference  frame   $K'$  is determined by
\begin{eqnarray}
x'(t') =  c  \; \cos \phi ' _{2} \; t '\; , \qquad z'(t') =  c \;
\sin \phi '_{2}  \; t '\; ; \label{2.1}
\end{eqnarray}

\noindent and the  parametric equation for the monopole string is
\begin{eqnarray}
x'  = L \; , \qquad z'  =  \lambda  \; , \;\; \lambda \in (-
\infty ,  + \infty ) \; . \label{2.2}
\end{eqnarray}

\noindent
For the  event  $2'$ one  should have two relations
\begin{eqnarray}
x_{2}' = c \; \cos \phi '_{2} \; t' _{2} = L \; , \qquad  z_{2}' =
c \;  \sin \phi '_{2}  \; t'_{2}  =  \lambda  \; \; . \label{2.3}
\end{eqnarray}

\noindent
Thus,   the coordinates of the event $2'$  in terms of $L, \phi'_{2}$ look
\begin{eqnarray}
ct'_{2} =    { L  \over \cos  \phi' _{2} } \; , \qquad x_{2}' =  c
\; \cos \phi '_{2} \; t' _{2}=   L \; ,  \qquad z'_{2}  = { L\;
\sin \phi ' _{2} \; \over \cos \phi'_{2}    } \; ; \label{2.4}
\end{eqnarray}

\noindent With the use of the Lorentz  formulas  one can establish
coordinates of the event $2'$ in the reference frame $K$ \footnote{Let $V = v/c$ and  $ct \Longrightarrow t$.}
\begin{eqnarray}
x_{2} = {x_{2}' -V \;t_{2}' \over \sqrt{1-V^{2}}} = { \cos
\phi_{2}'  - V \over \sqrt{1 - V^{2}}} \; t_{2}'  \; , \nonumber
\\
t_{2} = {t'_{2} - V \; x_{2}' \over \sqrt{1 -V^{2}}} = {1 - V \;
\cos \phi_{2}'  \over \sqrt{1 -V^{2}}} \; t_{2}' \; , \nonumber
\\
z_{2}= z_{2}'  = \sin \phi_{2} \; t_{2}' \; . \label{2.5}
\end{eqnarray}

\noindent Taking in mind
\begin{eqnarray}
t'_{2} = {\sqrt{1 - V^{2}} \over  1 - V \; \cos \phi_{2}'   } \;
t_{2} \; ,
\nonumber
%\label{2.6}
\end{eqnarray}

\noindent  one readily produces
\begin{eqnarray}
x_{2} = { \cos \phi_{2}'  - V \over 1 - V \; \cos \phi_{2}'} \; \;
t_{2}  \equiv \cos \phi_{2}  \; t_{2} \; , \nonumber
\\
z_{2}=  \sin \phi_{2}' \; \; {  \sqrt{1-V^{2}}   \over 1 - V \;
\cos \phi_{2}' }  \;\; t_{2} \equiv  \sin \phi_{2} \; t_{2}\; .
\label{2.7}
\end{eqnarray}

\noindent It is  easily verified that introducing the angle variable in  (\ref{2.7}) is correct
Indeed (in fact we  check below the invariance of the light velocity under Lorentz transformations)
\begin{eqnarray}
(\cos \phi_{2}) ^{2} + (\sin \phi_{2} )^{2} = ({ \cos \phi_{2}'  -
V \over 1 - V \; \cos \phi_{2}'} )^{2} + (\sin \phi_{2}' \; \; {
\sqrt{1-V^{2}}   \over 1 - V \; \cos \phi_{2}' })^{2} = \nonumber
\\
= {\cos^{2} \phi_{2}'  - 2 V  \cos \phi_{2}' + V^{2}  + \sin^{2}
\phi_{2}' - \sin^{2} \phi_{2}' V^{2} \over ( 1 - V \; \cos
\phi_{2}' )^{2} } = \nonumber
\\
= {1   - 2 V  \cos \phi_{2}'   + \cos^{2} \phi_{2}' V^{2} \over (
1 - V \; \cos \phi_{2}' )^{2} } = 1 \; . \nonumber
\end{eqnarray}

Let us write down the formulas describing relativistic aberration  effect (direct and inverse):
\begin{eqnarray}
 \cos \phi_{2} = { \cos \phi_{2}'  - V \over
1 - V \; \cos \phi_{2}'} \;  , \qquad \sin \phi_{2}= \sin
\phi_{2}' \; \; {  \sqrt{1-V^{2}}   \over 1 - V \; \cos \phi_{2}'
}   \; ,
% \label{2.8}
\nonumber
\end{eqnarray}
\begin{eqnarray}
 \cos \phi'_{2} = { \cos \phi_{2}  + V \over
1 + V \; \cos \phi_{2}} \;  , \qquad \sin \phi'_{2}= \sin \phi_{2}
\; \; {  \sqrt{1-V^{2}}   \over 1 + V \; \cos \phi_{2} }   \; .
\label{2.9}
\end{eqnarray}

\section{On geometrical form of the line of singularity \\  in the moving
reference  frame $K$}

\hspace{5mm} Let start with equation for the monopole string
in the rest reference frame  $K'$:
\begin{eqnarray}
x'  = L \; , \qquad z'  =  \lambda  \; , \qquad   \lambda \in (-
\infty , + \infty ) \; . \label{3.1}
\end{eqnarray}

To this geometrical line one may put in  correspondence
a special  set of events of the type $2' (2)$ in  space-time:
\begin{eqnarray}
t_{2}' =  {L  \over \cos  \phi'_{2} } \; , \qquad x_{2}' = \cos
\phi_{2}' \;\; t_{2}' \; , \qquad z_{2}' = \sin \phi_{2}' \;\;
t_{2}' \; . \label{3.2}
\end {eqnarray}

Space-time coordinate of these events  may be transformed to the moving reference frame $K$:
\begin{eqnarray}
t'_{2} = {t_{2} + V x_{2}  \over \sqrt{1 - V^{2} } } = {x_{2} /
\cos \phi _{2} + V x_{2}  \over \sqrt{1 - V^{2} } }  = {1 / \cos
\phi _{2} + V   \over \sqrt{1 - V^{2} } }\; x_{2} \;  , \nonumber
\\
x'_{2} = {x_{2} + V t_{2} \over \sqrt{1 - V^{2}}} = {x_{2} + V
x_{2} / \cos \phi_{2}  \over \sqrt{1 - V^{2}}} = {1 + V  / \cos
\phi_{2}  \over \sqrt{1 - V^{2}}}\; x_{2}   \; , \qquad z_{2}' =
z_{2} = \lambda    \; . \nonumber
\\
\label{3.3}
\end{eqnarray}

\noindent Expressing in (\ref{3.1})  the  variables in terms of new ones (not primed) we get
\begin{eqnarray}
{1 + V  / \cos \phi_{2}  \over \sqrt{1 - V^{2}}}\; x_{2}   = L ,
\qquad z = \lambda \label{3.4}
\end{eqnarray}

\noindent Now,  taking in mind
\begin{eqnarray}
x_{2} = \cos \phi_{2} \; t_{2} \; , \qquad z_{2} = \sin \phi_{2}
\; t_{2}: \nonumber
\\
 {1 \over \cos \phi_{2} } = \sqrt{1+ \tan^{2} \phi_{2} } =
\sqrt{1 + {z_{2}^{2} \over x_{2}^{2}}}  \; , \nonumber
\end{eqnarray}

\noindent eq. (\ref{3.4}) reads (index 2 is omitted)
\begin{eqnarray}
  {1 + V \sqrt{1 +  z^{2} /
 x^{2}}   \over \sqrt{1 - V^{2}}}\; \; x \; =  L , \qquad  z = \lambda  \in ( -\infty , + \infty ) \; ,
\label{3.5}
\end{eqnarray}

\noindent  or with the use of hyperbolical variable
\begin{eqnarray}
 ( \; \mbox{ch} \beta  + \mbox{sh} \beta \; \sqrt{1 +
z^{2} / x^{2}}\; ) \; x  = L \; . \label{3.6}
\end{eqnarray}

\noindent
Relations   (\ref{3.5})--(\ref{3.6})  should be  considered as describing the form of the monopole string
in the moving reference frame.

\hspace{5mm} Now, let us establish geometrical  forme of the curve given by
 (\ref{3.6}).  Eq.  (\ref{3.6}) reads
 \begin{eqnarray}
(L -  \mbox{ch} \beta \; x)^{2} =   \mbox{sh}^{2}\beta \; (x^{2} +
z^{2}) \; , \label{3.7}
\end{eqnarray}

\noindent and further
\begin{eqnarray}
L^{2}  - 2 L  x\mbox{ch} \beta \;   +  x ^{2} \;  =
 z ^{2} \; \mbox{sh}^{2} \beta \; ,\;\;
\nonumber
\end{eqnarray}

\noindent or
\begin{eqnarray}
( L^{2} - L^{2} \mbox{ch}^{2} \beta )+  (L \mbox{ch}  \beta  -  x)
^{2} \;  =
 z ^{2} \; \mbox{sh}^{2} \beta \; .
\nonumber
\end{eqnarray}

\noindent Evidently, this  curve is hyperbola:
\begin{eqnarray}
 {(L \mbox{ch}  \beta  -  x) ^{2}  \over L^{2} \mbox{sh} ^{2} \beta } \;  -  { z ^{2} \; \mbox{sh}^{2} \beta
 \over  L^{2} \mbox{sh} ^{2} \beta } = 1
\label{3.8}
\end{eqnarray}

\noindent There exists one special case: let  $L=0$, then latter equation  becomes equation for a direct line:
\begin{eqnarray}
  x ^{2}    -   z ^{2} \; \mbox{sh}^{2} \beta
  = 0 , \qquad x = \pm \; \mbox{sh} \beta \; z \; .
\label{3.9}
\end{eqnarray}

\section{
On the forme of the monopole string -- the case of arbitrary orientation of the string and velocity vector
 }

\hspace{5mm} Let  us  generalize the above result to the case of
arbitrary orientation of the monopole singular line and velocity vector.
We will need an explicit representation for Lorentz formulas with any velocity vector
 ${\bf V}$ (more details see in the book [12]). With the notation
 \begin{eqnarray}
{\bf V} = {\bf e} \; th\; \beta \;  , \qquad {\bf e} ^{2} = 1 \; ,
\nonumber
\\
{1 \over \sqrt{1 - V^{2} }} = ch\; \beta, \qquad {V \over \sqrt{1
- V^{2} }} = sh\; \beta, \qquad \label{4.1}
\end{eqnarray}

\noindent arbitrary Lorentz matrix  $L({\bf V}$ is
\begin{eqnarray}
(L_{a}^{\;\;b} )= \hspace{75mm} \nonumber
\\
\left | \begin{array}{cccc}
ch\; \beta       &   e_{1} sh\; \beta   &  e_{2} sh \;\beta  e_{2}  &  e_{3}sh\; \beta  \\
e_{1} sh\; \beta  & ch\; \beta -(ch\; \beta -1)(e_{2}^{2}
+e_{3}^{2} ) &
(ch\; \beta -1) e_{1} e_{2} &  (ch \; \beta -1) e_{2}e_{3}\\
 e_{2} sh\; \beta   & (ch\; \beta -1) e_{1}e_{2} &  ch\; \beta -(ch\; \beta -1)
 (e_{1}^{2} +e_{3}^{2} )&
(ch\; \beta -1) e_{2}e_{3} \\
 e_{3} sh\; \beta   & (ch\; \beta -1) e_{1}e_{3} & (ch\; \beta -1) e_{2}e_{3} &
ch\; \beta -(ch\; \beta -1)(e_{1}^{2} +e_{2}^{2} )\; ,
\end{array} \right |
\nonumber
\end{eqnarray}

\noindent  or differently
\begin{eqnarray}
(L_{a}^{\;\;b} )= \left | \begin{array}{cccc}
ch\; \beta       &   e_{1} sh\; \beta   &  e_{2} sh \;\beta  e_{2}  &  e_{3}sh\; \beta  \\
e_{1} sh\; \beta  &  1 + (ch\; \beta -1) e_{1}^{2} &
(ch\; \beta -1) e_{1} e_{2} &  (ch \; \beta -1) e_{2}e_{3}\\
 e_{2} sh\; \beta   & (ch\; \beta -1) e_{1}e_{2} &1 + (ch\; \beta -1) e_{2}^{2}  &
(ch\; \beta -1) e_{2}e_{3} \\
 e_{3} sh\; \beta   & (ch\; \beta -1) e_{1}e_{3} & (ch\; \beta -1) e_{2}e_{3} &
1 + (ch\; \beta -1) e_{3}^{2}
\end{array} \right | ,
\label{4.3}
\end{eqnarray}

\noindent  or in  symbolical form
\begin{eqnarray}
L = \left | \begin{array}{cc}
ch \; \beta & {\bf e}\;  sh\; \beta \\
{\bf e}  \; sh\; \beta & [\delta_{ij} + (ch\; \beta -1) e_{i}
e_{j}]\;
\end{array} \right | .
\label{4.4}
\end{eqnarray}

\vspace{5mm}
\noindent
Correspondingly, the Lorentz  transformation acts on space-time 4-vector $(t,{\bf x})$
as follows
\begin{eqnarray}
t' = ch\; \beta \; t +   sh\; \beta \; {\bf e}\;   {\bf x} \; ,
\nonumber
\\
{\bf x}'= {\bf e}  \; sh\; \beta  \; t + {\bf x}  + (ch\; \beta
-1)\; {\bf e} \; ({\bf e}  {\bf x}) . \label{4.5}
\end{eqnarray}

\noindent Inverse formulas are
\begin{eqnarray}
t = ch\; \beta \; t' -   sh\; \beta \; {\bf e}\;   {\bf x}'\; ,
\nonumber
\\
{\bf x} = -{\bf e}  \; sh\; \beta  \; t + {\bf x}'  + (ch\; \beta
-1)\; {\bf e} \; ({\bf e}  {\bf x}') \; . \label{4.6}
\end{eqnarray}

Let in the rest reference frame  $K'$  the monopole  string is given  by the parametric equation
\begin{eqnarray}
{\bf x}' =  \lambda \; {\bf n}  + {\bf x}_{0} \; . \label{4.7}
\end{eqnarray}

To this  geometrical line in 3-space  one  may put into correspondence a special set of space-time events
$\{ \; (t', {\bf x}' = (x'_{1},x_{2}',x_{3}') \;
\} $    defined as follows:
\begin{eqnarray}
{\bf x}' = {\bf c}' \; t', \qquad \Longrightarrow \qquad
 {\bf c}^{'2} = 1 \; , \qquad
t' = \sqrt{ {\bf x}^{'2}  } \; ; \label{4.8}
\end{eqnarray}

\noindent that is
\begin{eqnarray}
\left \{ \;\; t' = \sqrt{ {\bf x}^{'2}  } \; , \;
 {\bf x}' =  \lambda \; {\bf n}  + {\bf x}_{0}  \; \;  \right \} \; .
\label{4.9}
\end{eqnarray}

The same  events, being viewed from the  moving reference frame
 $K$, look as
\begin{eqnarray}
{\bf x} = {\bf c} \; t, \qquad \Longrightarrow \qquad
 {\bf c}^{2} = 1 \; , \qquad
t = \sqrt{ {\bf x}^{2}  } \; . \label{4.10}
\end{eqnarray}

Thus, the set of events (\ref{4.9}) now is described in term of new coordinates as follows:
\begin{eqnarray}
\left \{ \;\; t = \sqrt{ {\bf x}^{2}  } \; , \;
  {\bf x}  + {\bf e}  \;[ \;  sh\; \beta  \; t +
(ch\; \beta -1)\;  {\bf e}  {\bf x}\;  \; ]  = \lambda {\bf n} +
{\bf x} _{0} \; \;  \right \} \; . \label{4.10}
\end{eqnarray}

\noindent
Now, one should exclude the time-variable:
\begin{eqnarray}
  {\bf x}  + {\bf e}  \;[\;  sh\; \beta  \; \sqrt{ {\bf x}^{2}  } +
(ch\; \beta -1)\;  {\bf e}  {\bf x}\;  ]  = \lambda {\bf n} + {\bf
x} _{0}
 \; .
\label{4.11}
\end{eqnarray}

\begin{quotation}
{\em This,  we have produced equation describing the form the monopole string
in the moving reference  frame $K$. In the rest reference frame $K'$ (when  $\beta =0$)
eq. (\ref{4.11}) represent a direct line.}
\end{quotation}

In the first place, let us verify that the general equation  (\ref{4.11}) agrees with the particular
case studied above. Let it be
\begin{eqnarray}
{\bf n} = (0,0,1) \;, \qquad {\bf x}_{0}  = ( L, 0,0) \; , \qquad
{\bf e} = (1,0,0) \; , \nonumber
\end{eqnarray}

\noindent  then eq.  (\ref{4.11})  becomes
\begin{eqnarray}
x + \mbox{sh}\;  \beta \sqrt{{\bf x}^{2}} + ( \mbox{ch}\;  \beta
-1) x   = L \; , \qquad y = 0 , \qquad z =  \lambda \; ; \nonumber
\end{eqnarray}

\noindent which leads us to the above eq.  (\ref{3.7}):
\begin{eqnarray}
  \mbox{sh}^{2}  \beta \;  (x^{2}  +  z^{2} )  =  (L  - \mbox{ch}\; \beta )^{2} \; .
\nonumber
\end{eqnarray}

Now, consider a more general case:
\begin{eqnarray}
{\bf n} = (n_{1}, n_{2}, n_{3}) \; , \qquad {\bf e} = (1, 0,0 ) ,
\qquad {\bf x}_{0} = (0,0,0) \; . \label{4.12}
\end{eqnarray}

\noindent when a the singularity line goes through the origin $(0,0,0)$ of  $K'$.  At this eq.   (\ref{4.11} )
takes the form
 \begin{eqnarray}
\mbox{sh}\; \beta \sqrt{x^{2} + y^{2} + z^{2}}  +  x \;
\mbox{ch}\; \beta  = \lambda n_{1}  \; , \nonumber
\\
y = \lambda n_{2} , \qquad z =  \lambda n_{3} \;  . \label{4.13}
\end{eqnarray}

\noindent in the plane  $(y,z)$ one can introduce a rotated coordinate system $(Z,Y)$:
\begin{eqnarray}
z = Z  \; \cos \phi = Z \; { n_{3} \over \sqrt {n_{3}^{2} +
n_{2}^{2} }  } , \qquad y = Z \; \sin \phi  = Z \; { n_{2} \over
\sqrt{n_{3}^{2} + n_{2}^{2} }} \label{4.14}
\end{eqnarray}

\noindent  evidently  $\lambda$  looks as
\begin{eqnarray}
\lambda = {y \over  n_{2} } = { z \over n_{3}}  = \; { Z \over
\sqrt{n_{3}^{2} + n_{2}^{2} }} \; . \label{4.15}
\end{eqnarray}

\noindent Correspondingly, eq. (\ref{4.13})  gives
\begin{eqnarray}
\mbox{sh}\; \beta \sqrt{x^{2} + Z^{2} }  +  x \; \mbox{ch}\; \beta
= Z  \nu , \qquad \nu =
 { n_{1} \over \sqrt{n_{3}^{2} + n_{2}^{2} }}  = {n_{1} \over \sqrt{1 - n_{1}^{2}}}   \; .
\label{4.16}
\end{eqnarray}

\noindent From this it follows
\begin{eqnarray}
x^{2} \mbox{sh}^{2} \beta   +  Z^{2}  \mbox{sh}^{2} \beta       =
Z ^{2} \nu ^{2} -2  Z\nu \;  x \mbox{ch} \; \beta +
 x^{2} \; \mbox{ch}^{2}\; \beta  \; ,
\nonumber
\end{eqnarray}

\noindent or
\begin{eqnarray}
 ( \nu ^{2} - \mbox{sh}^{2}\beta ) \; Z ^{2}   -  ( \nu \mbox{ch}  \beta )  \;  2 Zx  + x^{2}   = 0  \; ,
\nonumber
\end{eqnarray}
that is
\begin{eqnarray}
 A = \nu ^{2} - \mbox{sh}^{2}\beta ,\;\;\;  B = \nu \mbox{ch}  \beta, \qquad A Z ^{2}   - B \;    2Zx  + x^{2}   = 0  \; .
\label{4.17}
\end{eqnarray}

\noindent Now, in the plane  $(x,Z)$ one can perform a special rotation $(x,Z) \Longrightarrow (X'',Z'')$
so that the  factor at $X'',Z''$-term  be equal to zero. Let it be
\begin{eqnarray}
x = \cos \phi X''  + \sin \phi Z'', \qquad Z = -\sin \phi X'' +
\cos \phi Z'' \; . \label{4.18}
\end{eqnarray}

\noindent then eq. (\ref{4.17})  becomes
\begin{eqnarray}
A\; [\;  \sin^{2} \phi   \; (X'')^{2}  - \sin 2\phi    \;  X''
Z'' +  \cos^{2}  \phi  \; (Z'')^{2}\; ] - \nonumber
\\
-  B\; [\; -\sin 2\phi  \; (X'')^{2}  + 2 \cos 2 \phi \;  X'' Z''
+ \sin 2\phi  \; (Z'')^{2} \; ] + \nonumber
\\
+ [ \;\cos^{2} \phi   \; (X'')^{2}  + \sin 2\phi    \;  X''  Z'' +
\sin^{2}  \phi  \; (Z'')^{2}\; ]=0 \; . \nonumber
\end{eqnarray}

\noindent From where, after re-grouping the terms we arrive at
\begin{eqnarray}
( A \;  \sin^{2} \phi + \cos^{2} \phi   + B\; \sin 2\phi       )
(X'')^{2} + \nonumber
\\
+
 ( A \cos^{2}  \phi  + \sin^{2} \phi  - B  \sin 2 \phi  ) \;  (Z'')^{2}
\nonumber
\\
-  [   ( A -1) \sin 2\phi   +2 B  \cos 2 \phi  ] X''  Z'' = 0\; .
\label{4.19}
\end{eqnarray}

\noindent Let us impose the restriction
\begin{eqnarray}
 ( A -1) \sin 2\phi   +2 B  \cos 2 \phi  = 0 , \qquad \Longrightarrow \qquad
\nonumber
\\
 \tan  2\phi =
  {2B \over 1 -A}  = {2  \nu \mbox{ch} \beta   \over   \mbox{ch}^{2}  \beta  - \nu^{2}    } =
 {2  \nu / \mbox{ch} \beta   \over   1 -   \nu^{2} / \mbox{ch}^{2}  \beta   } \; .
\nonumber
\end{eqnarray}

\noindent from which it follows
\begin{eqnarray}
\tan \phi = {\nu \over \mbox{ch} \beta } \; , \qquad \cos^{2} \phi
= {1 \over 1 + \tan^{2} \phi } = { \mbox{ch}^{2} \beta \over
\nu^{2} + \mbox{ch}^{2} \beta }\; , \qquad \sin^{2} \beta =  {
\nu^{2}  \over \nu^{2} + \mbox{ch}^{2} \beta } \; . \label{4.20}
\end{eqnarray}

\noindent
Therefore, eq. (\ref{4.19}) is much simplified:
\begin{eqnarray}
( A \;  \sin^{2} \phi + \cos^{2} \phi   + B\; \sin 2\phi       )
(X'')^{2} + \nonumber
\\
+
 ( A \cos^{2}  \phi  + \sin^{2} \phi  - B  \sin 2 \phi  ) \;  (Z'')^{2} = 0\;.
\nonumber
\end{eqnarray}

\noindent  Taking in  mid two identities:
\begin{eqnarray}
A \;  \sin^{2} \phi + \cos^{2} \phi   + B\; \sin 2\phi = {
(\nu^{2} - \mbox{sh}^{2} \beta )  \nu^{2}  +   \mbox{ch}^{2} \beta
+ 2 \nu ^{2} \mbox{ch}^{2}  \beta
 \over \nu^{2} + \mbox{ch}^{2} \beta } =
 \nonumber
 \\
 =
 { (\nu^{2} +1) ( \nu^{2} + \mbox{ch}^{2} \beta) \over \nu^{2} + \mbox{ch}^{2} \beta }=  (\nu^{2} +1) \; ,
\nonumber
\\
( A \cos^{2}  \phi  + \sin^{2} \phi  - B  \sin 2 \phi  ) = {
(\nu^{2} - \mbox{sh}^{2} \beta )  \mbox{ch}^{2} \beta +   \nu^{2}
- 2 \nu ^{2} \mbox{ch}^{2}  \beta
 \over \nu^{2} + \mbox{ch}^{2} \beta } =
\nonumber
\\
 = { \nu^{2} - \mbox{sh}^{2}  \beta \mbox{ch}^{2}  \beta -  \nu ^{2}  \mbox{ch}^{2} \beta  \over
 \nu^{2} + \mbox{ch}^{2} \beta  } =  -\mbox{sh}^{2} \beta \; ,
 \nonumber
 \end{eqnarray}

\noindent the latter relation  takes the form  of equation for a direct line:
\begin{eqnarray}
(\nu^{2} +1) (X'')^{2} - \mbox{sh}^{2} \beta (Z'')^{2} = 0 \; ;
\label{4.21}
\end{eqnarray}
remember  (see  (\ref{4.16}) )
\begin{eqnarray}\nu =
 { n_{1} \over \sqrt{n_{3}^{2} + n_{2}^{2} }}  = {n_{1} \over \sqrt{1 - n_{1}^{2}}}   \; .
\nonumber
\end{eqnarray}
It should be noted that in the case  $ n_{1}=0, n_{2}=0, n_{3} =1 $, the parameter  $\nu$
equals to zero, and eq.   (\ref{4.21})  reduces to (\ref{3.9}).

Now, consider  more general case (compare with  (\ref{4.11}))
\begin{eqnarray}
{\bf n} = (n_{1}, n_{2}, n_{3}) \; , \qquad {\bf e} = (1, 0,0 ) ,
\qquad {\bf x}_{0} = (L,0,0) \; ; \label{4.22}
\end{eqnarray}

\noindent Now, instead of (\ref{4.13})  we have
\begin{eqnarray}
\mbox{sh}\; \beta \sqrt{x^{2} + y^{2} + z^{2}}  +  x \;
\mbox{ch}\; \beta  = \lambda n_{1}  + L \; , \nonumber
\\
y = \lambda n_{2} , \qquad z =  \lambda n_{3} \;  . \label{4.23}
\end{eqnarray}

\noindent Further, one should act as above: in the plane  $(y,z)$
introduce the rotated system
\begin{eqnarray}
z = Z  \; \cos \phi = Z \; { n_{3} \over \sqrt {n_{3}^{2} +
n_{2}^{2} }  } , \qquad y = Z \; \sin \phi  = Z \; { n_{2} \over
\sqrt{n_{3}^{2} + n_{2}^{2} }} \nonumber
\\
\lambda = {y \over  n_{2} } = { z \over n_{3}}  = \; { Z \over
\sqrt{n_{3}^{2} + n_{2}^{2} }} \; , \qquad \nu =
 { n_{1} \over \sqrt{n_{3}^{2} + n_{2}^{2} }}  = {n_{1} \over \sqrt{1 - n_{1}^{2}}} \; ,
\nonumber
\end{eqnarray}

\noindent and from  (\ref{4.23})  it follows
\begin{eqnarray}
\mbox{sh}\; \beta \sqrt{x^{2} + Z^{2} }    = Z  \nu  -  x \;
\mbox{ch}\; \beta +L ,    \; . \label{4.24}
\end{eqnarray}

\noindent Further, instead of  (\ref{4.17}) we get
\begin{eqnarray}
 A = \nu ^{2} - \mbox{sh}^{2}\beta  \; ,\qquad B = \nu \mbox{ch}  \beta \; ,
 \nonumber
 \\
  A Z ^{2}   - B \;    2Zx  + x^{2}   + (L^{2} -2xL \mbox{ch} \beta  + 2Z  \nu L  )  = 0  \; .
\label{4.25}
\end{eqnarray}

\noindent Now, let us use the same rotation method:
\begin{eqnarray}
x = \cos \phi X''  + \sin \phi Z'', \qquad Z = -\sin \phi X'' +
\cos \phi Z'' \; , \nonumber
\end{eqnarray}
Eq. (\ref{4.25}) gives
\begin{eqnarray}( A \;  \sin^{2} \phi + \cos^{2} \phi   + B\; \sin 2\phi       ) (X'')^{2} +
 ( A \cos^{2}  \phi  + \sin^{2} \phi  - B  \sin 2 \phi  ) \;  (Z'')^{2}
\nonumber
\\
-  [   ( A -1) \sin 2\phi   +2 B  \cos 2 \phi  ] X''  Z''+
\hspace{40mm} \nonumber
\\
+
 [L^{2} -2   ( \cos \phi X''  + \sin \phi Z'' ) \; L \mbox{ch} \beta  +
 2 ( -\sin \phi X'' + \cos \phi Z''  ) \;   \nu L  ]  = 0\; ,
\label{4.26}
\end{eqnarray}

\noindent The angle we need is  already known:
\begin{eqnarray}
\tan \phi = {\nu \over \mbox{ch} \beta } \; , \qquad \cos^{2} \phi
= { \mbox{ch}^{2} \beta \over \nu^{2} + \mbox{ch}^{2} \beta }\; ,
\qquad \sin^{2} \beta =  { \nu^{2}  \over \nu^{2} + \mbox{ch}^{2}
\beta } \; . \nonumber
\end{eqnarray}

\noindent From  (\ref{4.25}) we get   (see (\ref{4.21}) )
\begin{eqnarray}
(\nu^{2} +1) (X'')^{2} - \mbox{sh}^{2} \beta (Z'')^{2} + \nonumber
\\
+ L^{2}  - 2 X'' L \; (   \mbox{ch}\beta  \cos \phi  + \nu  \sin
\phi ) - 2Z''  L \; (   \mbox{ch}\beta  \sin \phi  - \nu  \cos
\phi ) = 0 \; . \label{4.27}
\end{eqnarray}

\noindent Taking in mind identities
\begin{eqnarray}
   \mbox{ch}\beta  \cos \phi  + \nu  \sin \phi  =  { \mbox{ch}^{2} \beta  +
\nu ^{2} \over  \sqrt{\nu^{2} + \mbox{ch}^{2} \beta} } =
\sqrt{\nu^{2} + \mbox{ch}^{2} \beta} \; , \nonumber
\\
   \mbox{ch}\beta  \sin \phi  - \nu  \cos  \phi =  { \mbox{ch}\beta \; \; \nu - \nu \;  \; \mbox{ch}\beta  \over
\sqrt{\nu^{2} + \mbox{ch}^{2} \beta}} =0  \; , \qquad \nonumber
\end{eqnarray}

\noindent the previous equation reads as
\begin{eqnarray}
 (X'')^{2} - { \mbox{sh}^{2} \beta  \over (\nu^{2} +1)} (Z'')^{2} +
{L^{2} \over (\nu^{2} +1)}   - 2 X'' L \; { \sqrt{\nu^{2} +1  +
\mbox{sh}^{2} \beta} \over (\nu^{2} +1)} = 0 \; . \nonumber
\end{eqnarray}

\noindent It remains to perform an elementary  shift  in $X''$ variable:
\begin{eqnarray}
 \left [ X''   - L \; { \sqrt{\nu^{2} +1  +  \mbox{sh}^{2} \beta} \over (\nu^{2} +1)} \right ]^{2}
        -  { \mbox{sh}^{2} \beta  \over (\nu^{2} +1)} (Z'')^{2} =
 - {L^{2} \over (\nu^{2} +1)} +
        L ^{2} \; { \nu^{2} +1  +  \mbox{sh}^{2} \beta  \over (\nu^{2} +1)^{2} }
\nonumber
\end{eqnarray}

\noindent so that we  arrive at the equation determining  a hyperbola:
\begin{eqnarray}
 \left [ X''   - L \; { \sqrt{\nu^{2} +1  +  \mbox{sh}^{2} \beta} \over (\nu^{2} +1)} \right ]^{2}
        -  { \mbox{sh}^{2} \beta  \over (\nu^{2} +1)} (Z'')^{2} =
        { L ^{2} \mbox{sh}^{2} \beta  \over (\nu^{2} +1)^{2} }
\label{4.28}
\end{eqnarray}

\noindent In special case, when  $ n_{1}=0, n_{2}=0, n_{3}
=1 $, the parameter  $\nu$ becomes equal to zero, and
 (\ref{4.28})  is reduced to  (\ref{3.8}):
\begin{eqnarray}
 (x   - L  \mbox{сh} \beta )^{2}
        -   \mbox{sh}^{2} \beta  \;  z^{2} =
         L ^{2} \mbox{sh}^{2} \beta
\end{eqnarray}

Let us consider a more general case: let it be
\begin{eqnarray}
{\bf n} = (n_{1}, n_{2}, n_{3}) \; , \qquad {\bf e} = (1, 0,0 ) ,
\qquad {\bf x}_{0} = (x_{0} = L,y_{0}, z_{0}) \; ; \label{4.29}
\end{eqnarray}

\noindent at this we have (compare with (\ref{4.23}))
\begin{eqnarray}
\mbox{sh}\; \beta \sqrt{x^{2} + y^{2} + z^{2}}  +  x \;
\mbox{ch}\; \beta  = \lambda n_{1}  + L \; , \nonumber
\\
y = \lambda n_{2} + y_{0}, \qquad z =  \lambda n_{3}  + z_{0}\;  .
\label{4.30}
\end{eqnarray}

The  whole analysis of previous case is applied in full,
the only difference  arises  --  first one should perform a special shift
in the plane  $(y,z)$
\begin{eqnarray}
y   - y_{0} = \lambda n_{2} , \qquad z  - z_{0} =  \lambda n_{3}
\;  , \nonumber
\end{eqnarray}

\noindent  and then repeat the above calculation.

In conclusion, let us formulate general method to  modify  geometrical form of any rigid  curve (not only a direct line)
while changing the reference frame.
Let in the rest reference frame $K'$
 a curved line is  determined by two relations:
 \begin{eqnarray}
K':  \qquad \qquad \varphi_{1}({\bf x})  = 0  \; , \qquad
\varphi_{2}({\bf x}) = 0 \; . \label{A}
\end{eqnarray}

\noindent Evidently, the above receipt can apply  in this general case too --
it  gives two equations
\begin{eqnarray}
K:  \qquad \qquad \varphi_{1} \{ {\bf  {\bf x}  + {\bf e}  \;[\;
sh\; \beta  \; \sqrt{ {\bf x}^{2}  } + (ch\; \beta -1)\;  {\bf e}
{\bf x}\;  ]  \}} = 0 \; ,
\nonumber \\
  \varphi_{2} \{ {\bf  {\bf x}  + {\bf e}  \;[\;  sh\; \beta  \; \sqrt{ {\bf x}^{2}  } +
(ch\; \beta -1)\;  {\bf e}  {\bf x}\;  ]} \} = 0 \; . \label{B}
\end{eqnarray}

\noindent They  determine  geometrical form of the curved line in the moving reference frame.

\vspace{5mm}

{\em This  work was  supported  by Fund for Basic Research of Belarus and JINR  F06D-006.}

\end{document}